\documentclass[12pt]{article}

\usepackage{latexsym,amsmath,amssymb,amsthm,amsfonts,graphicx,placeins}
\usepackage{natbib,hyperref,chngcntr}
\usepackage{epsfig,bm,authblk,url,blkarray}

\usepackage{float} 
\usepackage{booktabs} 
\usepackage{graphicx} 
\usepackage[margin=1cm]{caption} 
\usepackage{subcaption}
\usepackage{mathtools}
\usepackage{xcolor}
\usepackage{color}
\floatstyle{plain}

\usepackage{algcompatible}
\usepackage{algorithmicx}
\usepackage[noend]{algpseudocode}
\usepackage{algorithm}
\usepackage{diagbox}
\newfloat{Algorithm}{thp}{lop}
\floatname{Algorithm}{Algorithm}

\usepackage{titling}
\DeclareMathOperator*{\argmax}{arg\,max}

\newcommand{\cov}{\text{Cov}}
\DeclareMathOperator{\ND}{\mathcal N}
\DeclareMathOperator{\UD}{U}
\DeclareMathOperator{\BetaD}{Beta}
\DeclareMathOperator{\logit}{logit}
\newcommand{\KLD}[2]{\text{KL}\left(#1\vert\vert#2\right)}

\begin{document}

\title{Logistic Gaussian process density regression:  a generalized Bayesian approach}
\date{\empty}
\author{
Zichuan Chen\thanks{\textit{Department of Statistics and Data Science, National University of Singapore}.},\,\,
Lucas Kock\thanks{\textit{Department of Statistics and Data Science, National University of Singapore}.},\,\,
Jeong Eun Lee\thanks{\textit{Department of Statistics, University of Auckland}.}\,\,
and
David J. Nott\thanks{Corresponding author:  standj@nus.edu.sg. \textit{Department of Statistics and Data Science, National University of Singapore} and \textit{Institute of Operations Research and Analytics, National University of Singapore}.}}

\maketitle
\vspace{-0.7in}

\begin{abstract}
Density regression extends conventional parametric regression by allowing 
the entire distribution of the response to vary flexibly with covariates rather than just low-order moments. In the 
Bayesian setting, logistic Gaussian process (GP) priors have been widely 
used for density estimation and extend naturally to density regression. 
The prior can be centred on a base density model, with the
nonparametric component providing an interpretable correction that is 
useful for model criticism. However, logistic GP density regression models 
have seen limited use, since they require computation of a normalizing 
constant for every observation, typically via numerical integration. 
We address this difficulty by proposing a generalized Bayesian
approach using a loss function based on the Hyv\"{a}rinen score. The Hyv\"{a}rinen 
score depends only on derivatives of the log density with respect to the 
response, eliminating the need to compute normalizing constants. 
Since GP computations remain expensive, we also employ sparse inducing 
point approximations and variational inference to develop a scalable approach. 
We demonstrate the method on one simulated and two real datasets, 
including a German weather dataset with more than 150,000 observations. 


\end{abstract}

\section{Introduction}\label{sec:Intro}

Bayesian methods for density regression have developed substantially in
recent years.  Conventional parametric regression models, where only low-order
moments are modelled as varying with covariates, may be inadequate to describe 
complex response-covariate relationships.  
Density regression overcomes this limitation by allowing the entire distribution 
to vary with covariates. We focus on density regression based on 
logistic Gaussian process (GP) priors and make three contributions.
 
First, logistic GP density regression methods lead to computational difficulties, 
since the likelihood terms contain observation-specific normalizing constants
that are unknown and are typically computed using numerical integration.  
This makes these regression density estimation methods hard to scale
to large datasets.  We address this by developing a generalized
Bayesian approach \citep{bissiri+hw16,knoblauch+jd22} 
where the loss-likelihood is based on the
Hyv\"{a}rinen score.  The Hyv\"{a}rinen score can be computed from
derivatives of the density with respect to the response, which does not
require knowledge of the normalizing constant.  
A second contribution of our
work is to develop a non-standard sparse inducing point approach for scalable variational
inference within this generalized Bayesian formulation.  Our third contribution
is to demonstrate the scalability of our methods through several examples, including one
large German weather dataset with over 150,000 observations.  Our examples also
demonstrate the value of the logistic GP approach for model criticism of
a parametric regression model. 

The logistic Gaussian process prior was first used for density 
estimation by \cite{leonard78}, and many authors have since developed 
it further in this setting, both technically and computationally, 
including \citet{lenk88,lenk91}, \cite{verdinelli+w88}, 
\cite{riihimaki+v14} and \citet{tokdar07} among many others.
A closely related prior is the Gaussian process density sampler
of \cite{murray+ma08}, with an improved MCMC implementation
provided by \cite{donner+o18}.  The work most relevant to ours
is \cite{paisley+zb25}.  The authors consider logistic Gaussian
process density estimation where the base density is Gaussian.  
For the Gaussian process they use a random Fourier feature approximation, 
and this allows them to derive several estimates of the density
in closed form.  The connection with our work lies in their use of a loss
function based on Fisher divergence in the estimation to avoid the computation of 
normalizing constants.  Our approach has no restriction on the base density 
and considers density regression and not just density estimation.

A common approach to selecting the loss function in generalized Bayes 
is to use an empirical estimate of some divergence \citep{jewson+sh18}. 
The Hyv\"{a}rinen 
score used here is
an empirical surrogate for an average Fisher divergence.  
Related discrepancies have been studied in generalized Bayesian
inference by many authors, motivated by either intractable
normalizing constants or robustness to model misspecification.   
\cite{matsubara+kbo22} consider kernel Stein discrepancies, including 
the Fisher divergence as a special case.  
The diffusion matching score discrepancy considered in \cite{barp+bdgm19} 
in the context of minimum discrepancy estimation is used in \cite{altamirano+bcess23} for 
Bayesian online changepoint detection 
where a weighting function is chosen to confer robustness. Similar 
approaches have been considered in  
\cite{altamirano+bk24} for robust estimation in Gaussian process 
regression, with extensions to spatio-temporal models 
\citep{laplante+adkb25} and multi-output Gaussian process 
regression \citep{rooijakkers+rbka25}.
\cite{bharti+dhb26} considers amortized simulation-based inference with 
exponential family models with intractable normalizing constants, 
although their approach can be used with other flexible density 
estimators as well.  \cite{jewson+r22} consider use of Hyv\"{a}rinen 
score for model selection between different loss functions in the 
generalized Bayesian framework. 

The logistic Gaussian process density regression model used here 
builds on the approach of \citet{tokdar+zg10}.  
They considered subspace projection in conjunction with their
density regression model to handle high-dimensional covariates.  
This extension could in principle be applied to our approach too, but
we do not pursue this here. The model of \citet{tokdar+zg10} 
develops further earlier work in \cite{tokdar07} on density estimation:  
both papers use a similar compactification 
of the Gaussian process domain via transformation 
and a low-rank approximation of the Gaussian process for 
computational tractability.  \citet{tokdar07} provides detailed guidance 
on node selection for the low-rank approximation.   
Some supporting theory is given in \citet{tokdar+g07}.
\citet{tokdar+zg10} use Markov chain Monte Carlo (MCMC) for posterior inference,
using numerical integration for computing the observation-specific normalizing 
constants appearing in the likelihood. 

An alternative approach 
to density regression with GPs is given by \cite{kundu+d14}, where the model
transforms latent uniform variables and the covariates while incorporating
an additive error term to define a flexible
distribution for the response.  The prior can be centred on a parametric
base model, similar to the logistic GP approach of \citet{tokdar+zg10}. 
With high-dimensional covariates, their model can be used in a 
latent factor formulation to achieve dimension reduction.
For computation, MCMC methods are used, and the unknowns include 
observation-specific latent variables.  This makes it hard to scale the
approach to large datasets; doing so 
would be an interesting direction for future work, but
we do not address it here.  
The most widely used approach to Bayesian density regression 
uses variations of the Dirichlet process (DP) \citep{ferguson73}.  These DP-based
methods are often, but not always, variations of
the dependent Dirichlet process \citep{maceachern99,maceachern00}.   
Commonly-used DP-based regression methods include those of 
\citet{deiorio+mrmm04}, \citet{griffin+s06} and \citet{dunson+pp07}, 
among many others.  
Here we focus on logistic GP priors in settings where criticism of
a base regression model is a focus.  For most DP-based methods, it is 
difficult to centre the nonparametric prior on a parametric regression model.  

In the next section, we introduce the logistic GP density regression model
considered throughout the rest of the paper.  Section 3 describes 
a generalized Bayesian inference approach for avoiding the computationally
expensive normalizing constant calculations in a conventional Bayesian
analysis.  Section 4 then considers variational approximations using
sparse inducing point methods, and our approach to estimating the learning
rate and other hyperparameters.  Several simulated and real examples
are discussed in Section 5, including a very large German weather dataset
with over 150,000 observations.  Section 6 gives concluding discussion. Code is publicly available at \url{https://github.com/DZCQs/GPDR}.

\section{Logistic GP density regression}\label{sec:Logistic}

We consider Gaussian process (GP) density regression in the formulation of 
\cite{tokdar+zg10}.  Let $x=(x_1,\dots, x_d)^\top$ be a $d$-dimensional covariate
and $y$ a scalar response.  To define the density regression
model of \cite{tokdar+zg10} we begin by transforming
$x$ to $w=S(x)=(S_1(x),\dots, S_d(x))$ in a compact domain.  
The definition of $S$ is example-dependent, and if $x_j$ already
lies in a compact interval we may take $S_j(x)=x_j$.  Specific 
choices of $S$ are discussed in our later examples.  
Next, we specify a base density model 
$g(y|x)$ for the response variable $y$ given $x$, with
corresponding distribution function $G(y|x)$.
Setting $z=G(y|x)$ maps $y$ to $z\in [0,1]$.
Considering a functional parameter $f(\cdot)$ on the space of $(w^\top,z)^\top$, 
the GP density regression model of \cite{tokdar+zg10} is 
\begin{align}
  h(y|x,f) & = g(y|x)\frac{\exp\left(f(S(x),G(y|x))\right)}{\int_0^1 \exp\left(f(S(x),z)\right)\,dz}, \label{rdgp}
\end{align}
with the function $f(\cdot)$ given a zero mean Gaussian process prior.  
If $f(\cdot)$ were identically zero, then $h(y|x,f)=g(y|x)$. In this
sense we can consider the nonparametric GP density regression model as ``centered''
on $g(y|x)$ and we can think of $f(\cdot)$ as defining a non-parametric correction to the base model $g(y|x)$. \cite{tokdar+zg10} observe that the transformation of $(x^\top,y)$ 
to variables $v=(w^\top,z)$ in a compact space ensures 
that the tails in the density regression model
behave appropriately. 
Theorem~3.1 in \cite{tokdar+zg10} guarantees posterior concentration when the base model $g(y\vert x)$ is heavy-tailed, and for
this reason we recommend using diffuse and heavy-tailed base models
in practice.

The Gaussian process prior for $f(\cdot)$ has mean zero and covariance function 
$\cov(f(v),f(v'))=\Delta(v,v')$.  The covariance function
$\Delta(\cdot,\cdot)$ will depend on additional lengthscale hyperparameters 
$\theta=(\theta_c^\top,\theta_r)$ where $\theta_c=(\theta_{c1},\dots, \theta_{cd})^\top\in \mathbb{R}^{d}$, $\theta_{cj}>0$, $j=1,\dots, d$ and $\theta_r> 0$.
Following \cite{tokdar+zg10}, we consider a Gaussian covariance function
\begin{align*}
\Delta(v,v') & = \sigma^2_f \exp\left(-\sum_{j=1}^d \theta_{cj}^2(w_j-w_j')^2-\theta_r(z-z')^2\right).
\end{align*}
For simplicity we do not show the dependence of $\Delta(\cdot,\cdot)$ on $\sigma^2_f$ and $\theta$
explicitly in our notation.

\section{Generalized Bayesian inference}\label{sec:Generalized}

One difficulty with working with the model \eqref{rdgp} is the need to approximate the 
normalizing constant $c(x,f)=\int_0^1 \exp(f(S(x),z))\,dz$ in computing the likelihood, 
which is a function of $x$.  
\cite{tokdar+zg10} use MCMC for computation, and approximate the Gaussian process
using a conditional expectation process given the values at a set of nodes, with 
normalizing constants computed using numerical integration.  Using numerical integration
on a grid for each observation likelihood term is slow and cumbersome, and here we 
suggest an alternative approach which does not require the normalizing constant, by using
a generalized Bayesian approach with the Hyv\"{a}rinen score \citep{Hyv2005} as the loss function.  

Suppose we have observed data $(x_i,y_i)$, $i=1,\dots, n$, where the $x_i$ are covariates
and the $y_i$ are corresponding responses.  Write $f_z(v)$ and $f_{zz}(v)$ for 
the first and second-order
partial derivatives of $f(v)$ with respect to $z$.  If $f(v)$ is a Gaussian process, then 
$(f(v),f_z(v),f_{zz}(v))^\top$ is a multivariate Gaussian process.  
In obvious notation we write $\Delta_{i,j}(v,v')$, $i,j\in \{0,1,2\}$ for the covariance between
the $i$th order partial derivative of $f(\cdot)$ with respect to $z$ at $v$ and the $j$th order 
partial derivative with respect to $z$ at $v'$ (for example,
$\Delta_{1,2}(v,v')=\text{Cov}(f_z(v),f_{zz}(v'))$).  We will use
$\Delta(v,v')$ and $\Delta_{0,0}(v,v')$ interchangeably.  
Let $w_i=S(x_i)$, $z_i=G(y_i|x_i)$ and
$v_i=(w_i^\top,z_i)^\top$.  We use the notation
$f=(f(v_1),\dots, f(v_n))^\top$, $f_z=(f_z(v_1),\dots, f_z(v_n))^\top$ and
$f_{zz}=(f_{zz}(v_1),\dots, f_{zz}(v_n))^\top$.   
Write
$\hbar(y|x,f):=\log h(y|x,f)$ and define
\begin{equation} 
  \mathcal{H}(y_i,x_i,f_z(v_i),f_{zz}(v_i)) := 2\hbar_{yy}(y_i|x_i,f)+\hbar_{y}(y_i|x_i,f)^2, \label{hscore}
\end{equation}
where $\hbar_y(y_i|x_i,f)$ is the first-order partial derivative of $\hbar(y_i|x_i,f)$ with respect to $y$, evaluated at $(y,x^\top)=(y_i,x_i^\top)$,
and $\hbar_{yy}(y_i|x_i,f)$ is the second-order partial derivative.  The right-hand side of \eqref{hscore}
depends on $f(\cdot)$ through $f_z(v_i)$ and $f_{zz}(v_i)$ only, and the expression is given in \ref{app:score}.  
Defining
\begin{align}
  \ell(y,x,f_z,f_{zz}) & = \frac{1}{n}\sum_{i=1}^n \mathcal{H}(y_i,x_i,f_z(v_i),f_{zz}(v_i)), \label{hscore2}
\end{align}
\eqref{hscore2} is the average Hyv\"{a}rinen score, which is an empirical
surrogate for the average Fisher divergence between $h(y|x,f)$ and the true conditional
density, denoted by $h_0(y|x)$, averaging over the distribution of $x$.  In generalized Bayesian inference \citep{bissiri+hw16,knoblauch+jd22} we replace a log-likelihood by a loss function in Bayes' rule, 
and using \eqref{hscore} as the loss, we obtain a generalized posterior density
\begin{align}
  \pi(f_z,f_{zz},\sigma^2_f,\theta|y) & \propto \pi(f_z,f_{zz}|\sigma^2_f,\theta)\pi(\sigma^2_f,\theta)
  \exp\left(-n \beta \ell(y,x,f_z,f_{zz})\right), \label{gpost1}
\end{align}
where $\beta>0$ is a learning rate.  
Here and elsewhere, 
we do not show the dependence of posterior distributions 
on the covariates or learning rate for simpler notation.
The connection between generalized Bayesian and conventional
Bayesian inference is the following.  If a negative log-likelihood
is chosen as the loss and the learning rate is 1, the generalized
Bayes posterior reduces to the conventional posterior.  However, when
using loss functions other than a log-likelihood as we do here, it is 
important to choose the learning rate so that the loss likelihood 
is properly scaled relative to the prior information.  There is a large
literature on the choice of learning rates in generalized Bayesian
inference (e.g. \cite{wu+m23}) and we discuss the choice of $\beta$ in
our approach in Section 4.  There are various ways to motivate
generalized Bayesian inference, such as considerations of coherence
under sequential updating \citep{bissiri+hw16} or frequentist justifications
in terms of PAC-Bayesian generalization bounds \citep{Alquier2024}.  
  
Since the loss function $\ell(y,x,f_z,f_{zz})$ depends only 
on $f_z$ and $f_{zz}$, and since the normalizing constant in 
$\log h(y|x,f)$ disappears when
we differentiate, generalized Bayesian inference based on \eqref{gpost1} avoids
calculation of normalizing constants which would be necessary when using conventional Bayesian 
inference.  Given the posterior inference on $f_z$ and $f_{zz}$, inferences on $f(v)$ for any $v$
could be obtained from the conditional prior density given $f_z$ and $f_{zz}$.  

Henceforth we will not consider a full Bayesian
treatment of the hyperparameters $\sigma^2_f$ and $\theta$, but select optimal point
estimates and the learning rate via cross-validation as described in \ref{appendix sec: hyperselection}.  We simplify notation 
by dropping dependence on $\sigma^2_f$, $\theta$ and $\beta$ for conditional
priors, posteriors and their approximations.  For example 
we write \eqref{gpost1} as
\begin{align}
  \pi(f_z,f_{zz}|y) & \propto \pi(f_z,f_{zz})\exp\left(-n \beta \ell(y,x,f_z,f_{zz})\right). \label{gpost}
\end{align}
We consider scalable variational methods for computing posterior approximations next.  

\section{Variational approximation for scalable computation}\label{sec:Variational}

To perform computations in a scalable way, we consider inducing variable methods 
(e.g. \citealp{titsias09}).  \cite{dezfouli+b15} and \cite{bonilla+kd19} consider inducing variable approaches to
computation in latent Gaussian process models which do not require any special structure
for the likelihood i.e.\ ``black box'' likelihoods.  This is ideal for applications of Gaussian process
models in generalized Bayesian inference.  Inducing point approaches 
augment the observed
inputs $v_1,\dots, v_n$, with additional inducing inputs $\widetilde{v}_1,\dots, \widetilde{v}_m$, 
and typically consider inducing values for $f$ at the inducing inputs.

Before describing our approach, we make some observations about 
the generalized posterior
\eqref{gpost} motivating the computational choices below.  
From \ref{app:score}, the Hyvärinen score $\mathcal{H}(y_i,x_i,f_z(v_i),f_{zz}(v_i))$
is quadratic in $f_z(v_i)$ and linear in $f_{zz}(v_i)$, and so the loss $\ell(y,x,f_z,f_{zz})$ in \eqref{hscore2} is a quadratic function of 
$(f_z,f_{zz})$. Since the prior for $(f_z, f_{zz})$ is Gaussian, the generalized posterior \eqref{gpost} is Gaussian.  
A generalized posterior obtained after introducing inducing variables
is also Gaussian.

One approach to exploiting this Gaussian structure would be to introduce inducing variables
for both $f_z$ and $f_{zz}$ and then use a single Gaussian variational approximation.  However, this doubles the number
of inducing values. Instead, we consider inducing values for the derivative
$f_z$ only, and replace $f_{zz}$ with an imputation.  This reduces the computational burden at the cost of an additional approximation.  

We write 
$\widetilde{f}_z=(f_z(\widetilde{v}_1),\dots, f_z(\widetilde{v}_m))^\top$.  
In explaining our algorithm 
the following notation will be used.   
Let $V$ be a matrix with $k$th row $v_k$, $k=1,\dots, n$, and $\widetilde{V}$ be a matrix with $k$th row
$\widetilde{v}_k$, $k=1,\dots, m$.  
Write $\Delta_{i,j}(V,\widetilde{V})$, $i,j\in \{0,1,2\}$, 
for the $n\times m$ matrix with $(k,l)$th entry $\Delta_{i,j}(v_k,\widetilde{v}_l)$.  
This notation is extended to other pairs of matrix inputs with rows in the space of $v$ by constructing
matrices with $(k,l)$th element evaluating $\Delta_{i,j}(\cdot,\cdot)$ at the $k$th row of the first
matrix argument and $l$th element of the second matrix argument.    

We replace the values $f_{zz}$ with an imputation 
$\widehat{f}_{zz}=(\widehat{f}_{zz}(v_1),\dots, \widehat{f}_{zz}(v_n))^\top$ which is given by 
\begin{equation}\label{eq:hatf_zz}
    \widehat{f}_{zz}=E\left[f_{zz}|\widetilde{f}_z\right]=\Delta_{2,1}(V,\widetilde{V})\Delta_{1,1}(\widetilde{V},\widetilde{V})^{-1}\widetilde{f}_z.
\end{equation}
We then infer about $f_z$ using the augmented generalized posterior density including the inducing values,  
\begin{align}\label{eq:gen_post}
 \pi(\widetilde{f}_{z},f_z|y) & \propto 
 \pi(\widetilde{f}_z)\pi(f_z|\widetilde{f}_z)
 \exp\left(-\beta \sum_{i=1}^n \mathcal{H}(y_i,x_i,f_z(v_i),\widehat{f}_{zz}(v_i))\right),
\end{align}
and a structured variational approximation of the posterior of the form
\begin{align*}
  q(\widetilde{f}_z,f_z) & = q(\widetilde{f}_z)\pi(f_z|\widetilde{f}_z).
\end{align*}
After substituting the imputation \eqref{eq:hatf_zz} 
the target \eqref{eq:gen_post}
is Gaussian in $(\widetilde{f}_z, f_z)$.  In the variational approximation, the conditional prior
of $f_z|\widetilde{f}_z$ is used as the conditional variational posterior
density of $f_z|\widetilde{f}_{z}$, which simplifies computation, as is standard
with inducing point approaches to GP computation.    

Although the generalized posterior \eqref{eq:gen_post} is Gaussian,  
we use a diagonal mixture of Gaussians as the variational approximation
for the inducing values.  For a small number
of mixture components, the mixing distribution
allows capturing dependence structure, but the
diagonal covariance matrices for the components
reduces the number of variational parameters
compared to using a single Gaussian approximation
with a dense covariance matrix.
Specifically, we assume
$$q(\widetilde{f}_z)=\sum_{k=1}^C \omega_k q_{k}(\widetilde{f}_z),$$
where $C$ is the number of mixture components, $\omega_k\geq 0$ are mixing weights, 
$k=1,\cdots, C$, $\sum_{k=1}^C \omega_k=1$ and
$q_{k}(\widetilde{f}_z)$ is the $k$th Gaussian mixture component, 
$$q_k(\widetilde{f}_z)=\phi(\widetilde{f}_z;\mu_{k},\Sigma_{k}),$$
where $\phi(x;\mu,\Sigma)$ denotes a multivariate normal distribution with mean vector $\mu$
and covariance matrix $\Sigma$.
We restrict the covariance matrices to be diagonal with diagonal elements $\sigma^2_{kj}$, $k=1,\dots,C$, $j=1,\dots,m$, and set $\omega_k=\frac{\exp(w_k)}{1+\sum_{k=1}^{C-1}\exp(w_k)}$ for $k=1,\dots,C-1$ and $\omega_C=\frac{1}{1+\sum_{k=1}^{C-1}\exp(w_k)}$, so that $$\lambda=(\mu_1^\top,\dots,\mu_C^\top,\log(\sigma^2_{11}),\dots,\log(\sigma^2_{1m}),\log(\sigma^2_{21}),\dots,\log(\sigma^2_{Cm}),w_1,\dots,w_{C-1})^\top$$ is the vector of unconstrained variational parameters to be optimized.
In what follows 
we do not show the dependence of variational densities on
$\lambda$ explicitly for simpler notation. 
We select inducing point locations by 
applying $k$-means clustering to the training 
data taking the cluster centroids as the inducing
points, although it is also possible to optimize 
the inducing points.

Following \cite{dezfouli+b15}, the optimization target for $\lambda$ is given by 
\begin{equation}\label{opt obj}
    \mathcal{L}(\lambda)=-\KLD{q(\widetilde{f_z})}{\pi(\widetilde{f_z})}-\beta E_q\left[\sum_{i=1}^n \mathcal{H}(y_i,x_i,f_z(v_i),\widehat{f}_{zz}(v_i))\right],
\end{equation}
where $\widehat{f}_{zz}$ in the loss is evaluated at the current value of $\lambda$ through Equation (\ref{eq:hatf_zz}).
The Hyvärinen score is quadratic in $\widetilde{f_z}$ under the approximation \eqref{eq:hatf_zz},
which makes the expectation in \eqref{opt obj} tractable.
We approximate the KL-divergence term using Goldberger's approximation \citep{hershey2007approximating,GolGorGre2003},
\begin{equation}\label{eq:goldberg_approx}
    \KLD{\sum_{k=1}^C \omega_k q_k(\widetilde f_z)}{\pi(\widetilde f_z)}
\approx
\sum_{k=1}^C \omega_k
\KLD{q_k(\widetilde f_z)}
{\pi(\widetilde f_z)} + \sum_{k=1}^C\omega_k \log (\omega_k),
\end{equation}
where on the right-hand side the KL-divergence terms between Gaussian distributions appearing in the first sum have closed form expressions, while the second sum acts as an additional entropy correction for the mixture weights. We use standard stochastic gradient ascent with the adaptive ADAM learning rate \citep{KinBa2014} to derive the optimal variational parameters $\lambda^*=\argmax\mathcal{L}(\lambda)$,
using automatic differentiation for computation
of gradient estimates. Further details are given in \ref{app:klbound}.

\subsection{Computation of predictive densities}

For predictive inference, suppose we want to plot an estimate of the predictive density of the response
given $x=\breve{x}$ for some $\breve{x}$. We consider an equally spaced 
grid of points with grid spacing $\epsilon$ for $y$, $\breve{y}_1<\dots <\breve{y}_M$.  
Corresponding $z$ values $\breve{z}_j$ are defined as $G(\breve{y}_j|\breve{x})$, $j=1,\dots, M$, and 
we write $\breve{v}_j=(S(\breve{x})^\top,\breve{z}_j)^\top$.  
Write $\breve{V}$ for the matrix with $j$th row $\breve{v}_j$, 
and $\breve{f}=(f(\breve{v}_1),\dots, f(\breve{v}_M))^\top$. 

An approximate posterior mean \(\overline f\) of \(\breve f\) is obtained by averaging the conditional GP prior mean \(E_{\pi}(\breve f\mid\widetilde f_z)\) with respect to the variational distribution \(q(\widetilde f_z)\). Because this conditional mean is linear in \(\widetilde f_z\), the resulting expression is equivalent to evaluating it at \(E_q(\widetilde f_z)\):


$$\overline{f}\approx \Delta_{0,1}(\breve{V},\widetilde{V})\Delta_{1,1}(\widetilde{V},\widetilde{V})^{-1}\left(\sum_{k=1}^C \omega_k\mu_k\right).$$
Write $\overline{f}(\breve{v_j})$ for the $j$th element of $\overline{f}$. Using the Riemann-sum approximation to the normalizing constant, $h(\breve{y}_j|\breve{x},f)$, $j=1,\dots, M$ is estimated as
$$\widehat{h}(\breve{y}_j|\breve{x},f)= g(\breve{y}_j|\breve{x})\frac{\exp(\overline{f}(\breve{v_j}))}
{\sum_{k=2}^M \exp(\overline{f}(\breve{v_j}))(\breve{z}_k-\breve{z}_{k-1})}.$$
 
\section{Simulation studies}\label{sec:Examples}

This section illustrates our Gaussian process density regression (GPDR) 
approach on one simulated and two real datasets. Before presenting these 
examples, we discuss the choice of base measure $g(y|x)$ and how to 
interpret the nonparametric correction $f(\cdot)$ that GPDR provides.
As noted in Section~2, when $f(\cdot)$ is identically zero, $h(y|x,f)=g(y|x)$, 
so $f(\cdot)$ represents the correction that the nonparametric model applies 
to the base measure. Specifically, rearranging~\eqref{rdgp}, we have 
\begin{align}
    \log h(y|x,f) = \log g(y|x) + f(S(x),G(y|x)) + C(x,f), \label{logrdgp}
\end{align}
where $C(x,f)=-\log\int_0^1\exp(f(S(x),z))\,dz$ does not depend on $y$. 
Thus, the shape of $f(S(x),z)$ directly 
describes how the nonparametric model corrects the log-density of the 
base measure.

However, as discussed in Theorem~3.1 of \cite{tokdar+zg10}, it is 
important that the base measure $g(y|x)$ has sufficiently heavy tails 
to ensure posterior concentration. For this reason, when our scientific 
interest lies in criticising a parametric model $g_0(y|x)$, we recommend 
constructing $g(y|x)$ as a more diffuse and heavy-tailed version of 
$g_0(y|x)$ rather than setting $g(y|x)=g_0(y|x)$. To understand the correction that the GPDR model implies 
for $g_0(y|x)$, we note that
\begin{align}
    \log h(y|x,f) - \log g_0(y|x) = \log\frac{g(y|x)}{g_0(y|x)} 
    + f(S(x),G(y|x)) + C(x,f), \label{correction}
\end{align}
where $C(x,f)$ does not depend on $y$. Hence the correction $f_0(S(x),y):=\log\frac{g(y|x)}{g_0(y|x)} 
    + f(S(x),G(y|x))$ to 
$g_0(y|x)$ combines the log-ratio 
$\log g(y|x)/g_0(y|x)$ with the GP correction term $f(S(x),z)$. 
Specific choices of $g_0(y|x)$ and $g(y|x)$ are given in the 
examples below.  We can also write 
\begin{align}
  h(y|x,f) & \propto g_0(y|x)\exp(f_0(S(x),y)),  \label{f0}
\end{align}
and we can think of the original model with base density model $g(y|x)$ as corresponding
to a similar model correcting the parametric model $g_0(y|x)$ but with the Gaussian
process correction $f_0(S(x),y)$ no longer having zero mean.  The criticism of a specified
parametric model is the focus of the three examples below.  

Throughout, we compare our GPDR approach to established methods from the distributional regression literature. We consider a parametric generalized additive model for location, scale, and shape \citep[GAMLSS;~][]{RigSta2005}, in which all parameters of a parametric response distribution are linked to structured additive predictors of the covariate $x$; a conditional transformation model \citep[TRAM;~][]{HotKneBue2014}, which learns a semi-parametric, covariate-dependent transformation of a Gaussian base measure to model the response distribution $y\vert x$; a quantile regression forest \citep[QRF;~][]{AthTibWag2019}, which uses adaptive, forest-derived nearest-neighbor weights to estimate the conditional response distribution non-parametrically; and a Bayesian non-parametric covariate dependent Dirichlet process mixture \citep[DDPstar;~][]{RodInaKle2025}. An empirical comparison with the MCMC sampling approach for logistic Gaussian process density regression of \citet{tokdar+zg10} was not feasible, as there is no existing available implementation for this method.

We illustrate the proposed approach through three examples: a simulation study based on Gaussian regression in Section~\ref{subsec:simulation} and two real-data applications concerning spatio-temporal weather dynamics in Section~\ref{subsec: weather} and the Gini index with a non-Gaussian response in Section~\ref{subsec: gini}). The hyperparameter estimates for GPDR in each study are summarized in
Table~\ref{tab:hyper_sim}. To assess the predictive performance of GPDR in comparison to the benchmarks, we report the average log-score, the RMSE, $\sqrt{n^{-1}\sum_{i=1}^{n}(y_i-\mathbb{E}[Y_i\mid x_i])^2}$, the empirical coverage of equal-tailed $95\%$ prediction intervals, and their average length. These results are summarized in Table~\ref{tab:metrics}. All performance metrics are evaluated on the hold-out test dataset. 

\subsection{Gaussian regression}\label{subsec:simulation}
As a first example, we consider a deliberately misspecified regression model. Data $(y_i,x_i)$, $i=1,\dots,n$, are generated as 
\begin{align*}
    x&\sim\UD(0,1)\\
    y\mid x&\sim \ND(x+x^2,(0.2x+0.05)^2),
\end{align*}
and then a linear Gaussian regression model is fitted,
\begin{align*}
    y = bx+\sigma\varepsilon,
\end{align*}
where $\varepsilon\sim\ND(0,1)$, and $\eta=(b,\sigma)^\top$ are model parameters. This parametric model is misspecified as it cannot capture the non-linear relationship between $y$ and $x$ in the mean nor the heteroscedastic noise. Since $x\in[0,1]$, we do not consider any transformation of $x$ and work with $\nu=(x,z)^\top$ directly.  The model parameters are estimated below from the training data to obtain estimates $(\hat{b},\hat{\sigma})$ of $(b,\sigma)$.  From
these estimates we obtain plug-in predictive densities 
for the parametric model, $g_0(y|x)$.

Since the parametric model is not heavy-tailed, we consider the more diffuse model where $y\vert x\sim t_3(\hat{b}x,(3\hat{\sigma})^2)$ as the base density model $g(y|x)$ for training GPDR, where $t_\xi(\cdot)$ denotes a Student-$t$ distribution with $\xi$ degrees of freedom. This
diffuse and heavy-tailed $g(y|x)$ avoids placing many transformed responses $z=G(y\mid x)$ close to the boundaries of $[0,1]$.  
If there are many values of $z$ near the boundary, to correct
the base density we need the Gaussian process correction to change
rapidly at the boundary, which is difficult when a stationary
Gaussian process prior is used for $f(\cdot)$.  

\begin{figure}[tbh]
\centering
\begin{subfigure}{0.31\linewidth}
         \includegraphics[width=\linewidth]{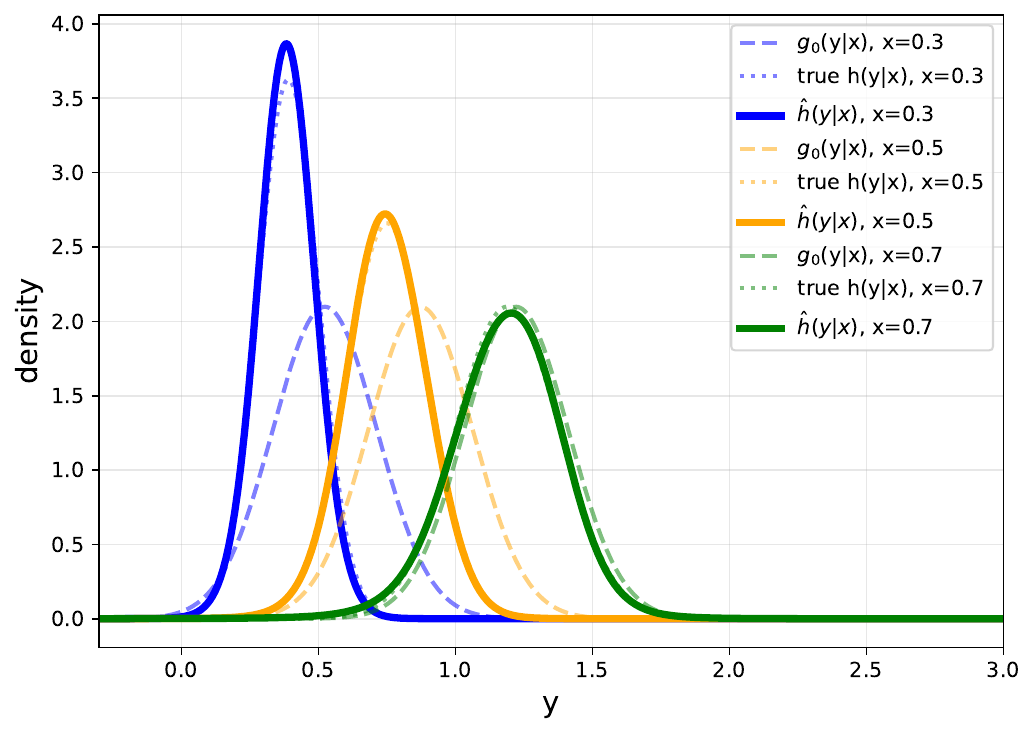}
         \caption{Predictive densities}
         \label{fig1: panel A}
\end{subfigure}
\hfill
\begin{subfigure}{0.31\linewidth}
         \includegraphics[width=\linewidth]{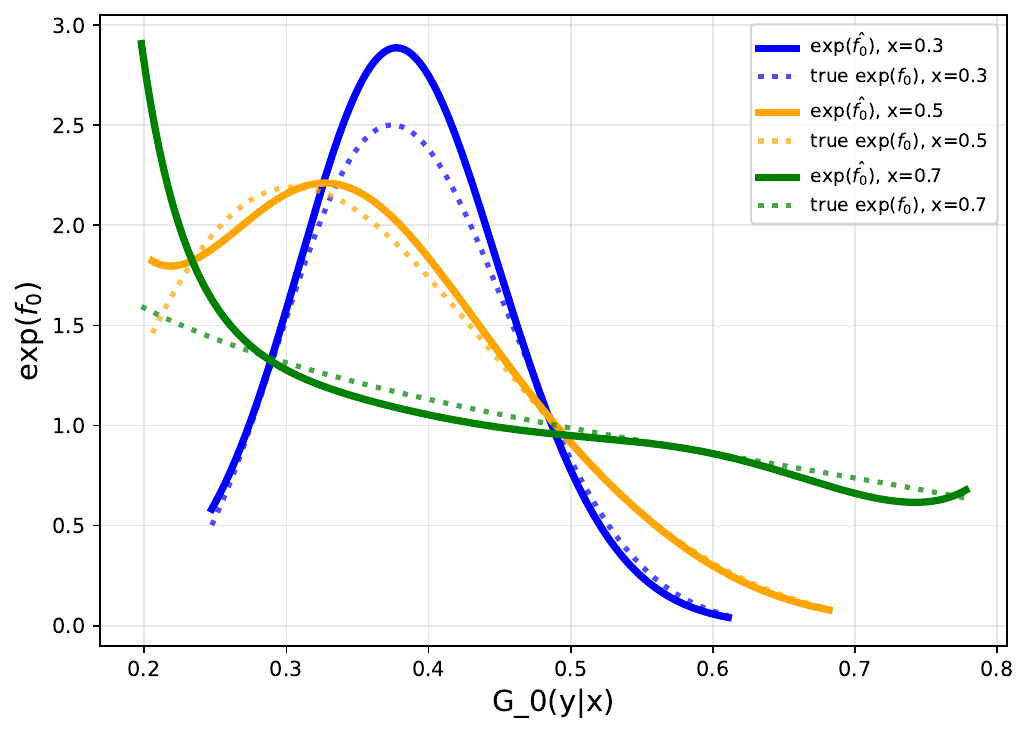}
         \caption{$\exp(f_0(x,y))$}
         \label{fig1: panel B}
\end{subfigure}
\hfill
\begin{subfigure}{0.31\linewidth}
         \includegraphics[width=\linewidth]{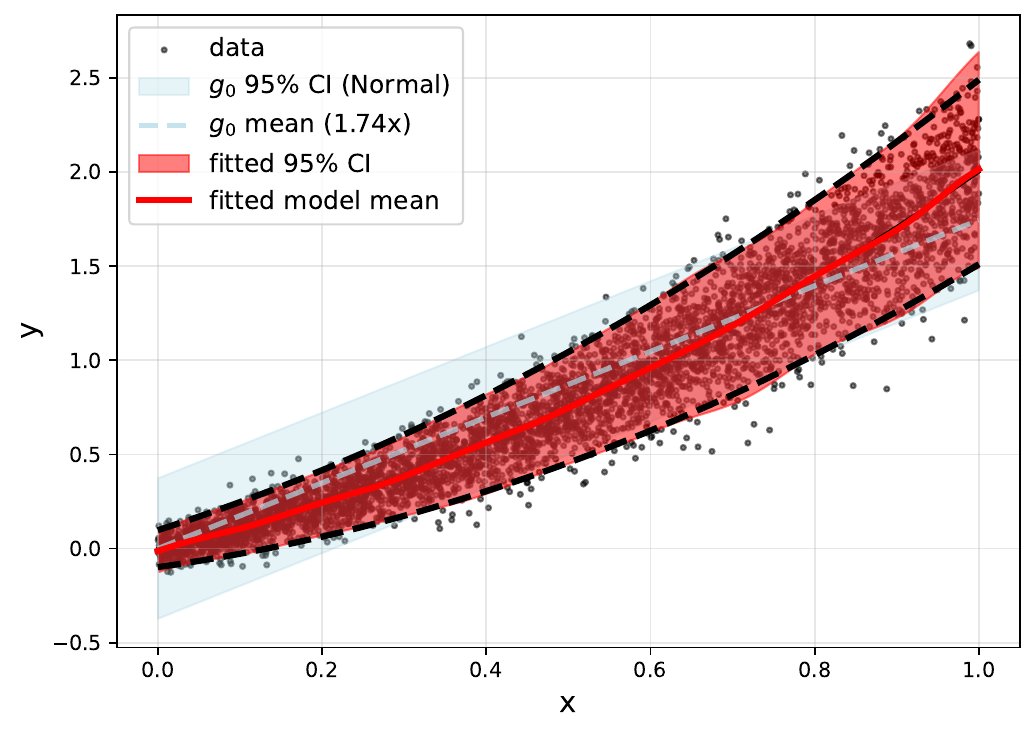}
        \caption{Regression fit}
         \label{fig1: panel C}
\end{subfigure}

\caption{\small Simulations. \textbf{a)} Predictive densities at $x=0.3$ (blue), $x=0.5$ (orange), and $x=0.7$ (green) for GPDR (bold), the true DGP (dotted), and the Gaussian parametric model (dashed). 
The dotted lines overlap with GPDR.  \textbf{b)} shows $\exp(f_0)$ corresponding to the densities in \textbf{a)} respectively. We show the theoretical optimum up to a constant under which GPDR would match exactly with the true DGP (dotted) in terms of these estimated functions (bold). \textbf{c)} Comparison between the base model (light blue dashed) and GPDR (red bold). We show the conditional mean, $\mathbb{E}[y\vert x]$, as well as equal tailed $95\%$ prediction intervals (shaded). The black-dashed lines show 95\% equal tailed intervals for the true DGP.}
\label{fig:toy}
\end{figure}

We generate $n_\text{train}=5,000$ observations for training of GPDR, and $n_\text{test}=50,000$ observations from the true data generating process (DGP) for model evaluation. In Table~\ref{tab:metrics} GPDR substantially improves on the misspecified parametric base model performing on par with GAMLSS, the best performing benchmark on this example. The good performance of GAMLSS is expected here, as GAMLSS assumes the correct parametric form of the data generating process. This is substantial additional information that is not available to the other benchmark methods, including GPDR. The empirical coverage of the 95\% credible interval derived from GPDR is close to the nominal level of $0.95$, indicating a well-calibrated model that captures the full conditional response distribution rather than the mean function alone. This conclusion is further supported by the log-score. 

Figure~\ref{fig1: panel A} compares the predictive densities from the parametric model, $g_0(y\vert x)$, and under GPDR, $h(y\vert x,f)$, with the true conditional densities from the DGP at $x=0.3, 0.5, 0.7$. 
Again, GPDR greatly improves the parametric model and closely matches the true DGP even in areas of severe misspecification.  
Figure~\ref{fig1: panel B} shows $\exp(f_0(x,y))$ as a function of $z_0=G_0(y|x)$ for $x=0.3, 0.5, 0.7$. Recall that $f_0(x,y)$ in \eqref{f0} can be interpreted as the correction to the parametric density model $g_0(y|x)$.  The plot of $\exp(f_0(x,y))$ is helpful for model criticism as the estimated functions are highly interpretable. For a given value $x$, the shape of $f_0(x,y)$ can be interpreted in a similar manner to the rank histogram inspection proposed by \citet{GneBalRaf2007}. The parametric model $g_0(y\vert x)$ is biased for $x=0.3, 0.5, 0.7$, which is indicated by the skewness of $\exp (f_0(x,y))$. The triangle shapes of $\exp(f_0(x=0.3,y))$ and $\exp(f_0(x=0.5,y))$ indicate overdispersion as the base model overestimates uncertainty at these locations. $\exp(f_0(x=0.7,y))$ is much closer to a constant function as $g(y\vert x=0.7)$ matches the true DGP more closely and deviation is most pronounced in the upper tail. 

Figure~\ref{fig1: panel C} shows the $n_\text{train}$ samples from the true data generating process along with the predictive means and $95\%$ prediction intervals from the parametric model $g_0(y\vert x)$ and GPDR $h(y\vert x,f)$. The parametric model produces biased predictions and many observations lie outside the $95\%$ prediction interval. GPDR corrects the predictive mean by aligning it closely with the true mean $x+x^2$ from the DGP and improves uncertainty quantification by correcting for heteroskedasticity. Although $g_0(y|x)$ fits the data poorly, GPDR closely matches the true DGP.  Please refer to Table 2 for hyperparameter estimates used for this
example.

\subsection{Spatio-temporal weather dynamics}\label{subsec: weather}

The German Weather Service (Deutscher Wetterdienst; DWD) provides meteorological measurements from a large number of weather stations across Germany. Here, we analyze the daily maximum temperature observed at a total of 463 stations in 2020 
constituting a large data set with a total of $n=167,762$ observations of which we use a random subset of $n_\text{train}=150,985$ observations for training and the remaining $n_\text{test}=16,777$ observations for model evaluation. Training a Gaussian process on a dataset of this size is challenging, and this example illustrates how the inducing point approach allows for scalability of GPDR. 

For observation $i=1,\dots,n$, let $y_i$ denote the maximum temperature, $\texttt{day}_i$ the day of the year, and $(\texttt{longitude}_i,\texttt{latitude}_i)$ the coordinates of the corresponding weather station, so that $x_i=(\texttt{day}_i,\texttt{longitude}_i,\texttt{latitude}_i)^\top$ denotes the covariate vector. To understand the spatio-temporal dynamics, we consider an additive model
\begin{align*}
    y_i = \beta_0+\mathfrak{f}_\text{temp}(\texttt{day}_i)+\mathfrak{f}_\text{spat}(\texttt{longitude}_i,\texttt{latitude}_i)+\varepsilon_i,
\end{align*}
where $\beta_0$ is an intercept, $\mathfrak{f}_\text{temp}(\cdot)$ is a smooth non-linear effect of time, $\mathfrak{f}_\text{spat}(\cdot,\cdot)$ is a smooth bivariate spatial effect, and $\varepsilon_i\sim\ND(0,\sigma^2)$ is iid noise. Both non-linear effects are represented through a spline basis and the additive model is fitted via constrained maximum likelihood estimation as implemented in the python package pyGAM \citep{SerBru2018}. The fitted effects shown in \ref{appendix sec: pygam on weather} capture expected dynamics: higher temperatures in summer compared with winter, and temperatures in northern Germany close to the sea or at the hillside in south-east Germany are expected to be higher than in the remaining regions.  We will use our 
nonparametric GPDR approach to criticize plug-in predictive densities 
$g_0(y|x)$ from the Gaussian additive model.

Before training GPDR, we standardize all covariates and map them to $[-1,1]$ using a sigmoid transformation. As in Example~\ref{subsec:simulation}, we consider diffuse and heavy-tailed base densities $t_3(\beta_0+\mathfrak{f}_\text{temp}(\texttt{day})+\mathfrak{f}_\text{spat}(\texttt{longitude},\texttt{latitude}),(3\sigma)^2)$ for $g(y|x)$, where 
we plug in estimates from the Gaussian model for the unknown parameters and functional terms.  
The additive Gaussian model cannot capture joint dynamics of time and space and assumes a simple homoscedastic error model, which might be too limiting. Consequently, GPDR results in a substantial improvement compared to the base model in terms of log-score and RMSE evaluated on the hold-out test set as shown in Table~\ref{tab:metrics}. GPDR outperforms all other benchmarks in this example. 

\begin{figure}[ht!]
    \centering
    \includegraphics[width=0.9\linewidth]{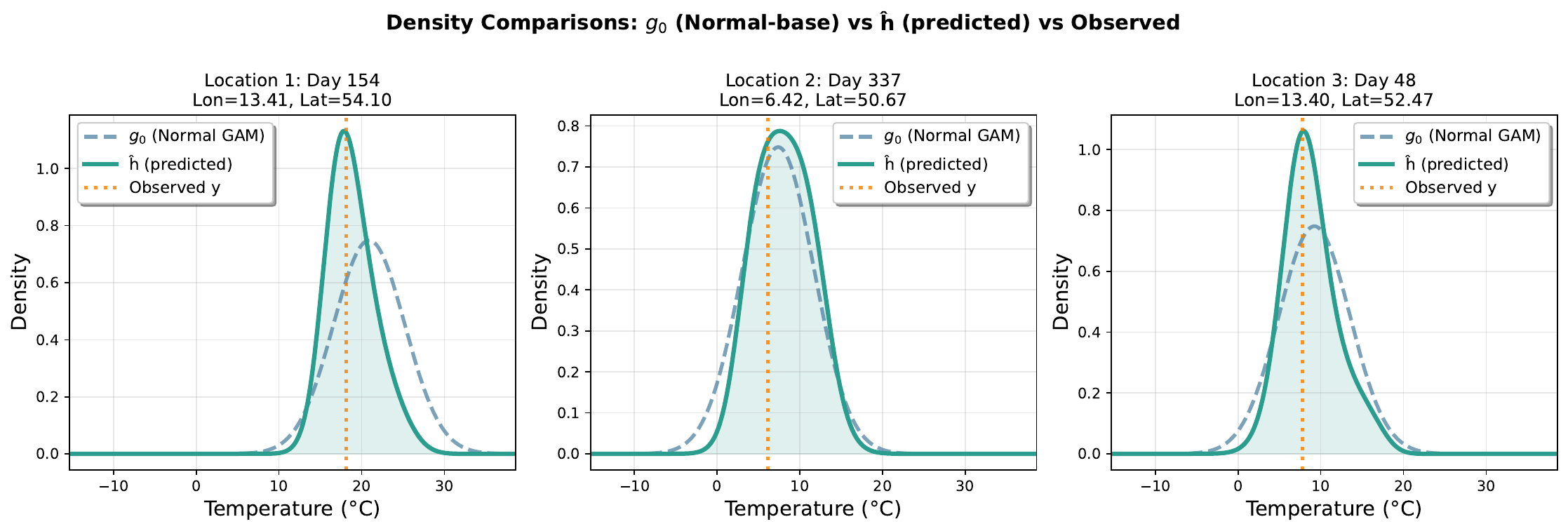}
    \caption{\small GPDR Predictive density (solid), $g_0(y|x)$ (dashed) and observed data $y$ (dotted vertical line) at three randomly selected time-location pairs. From left to right the predicted density is evaluated at $(\texttt{day},\texttt{longitude},\texttt{latitude}) = (154, 13.41, 54.10), (337, 6.42, 50.67), (48, 13.40, 52.47)$, respectively.}
    \label{fig:weather_density}
\end{figure}

Figure~\ref{fig:weather_density} shows the predictive densities at three representative time–location pairs. 
Compared with $g_0(y|x)$, the GPDR predictive densities place their modes closer to the observed responses while  reducing predictive uncertainty. This behavior is consistent with the results reported in Table~\ref{tab:metrics}. The predictive densities are also non-Gaussian, showing skewness and thus substantially deviating from  $g_0(y|x)$.

\begin{figure}
    \centering
    \includegraphics[width=0.9\linewidth]{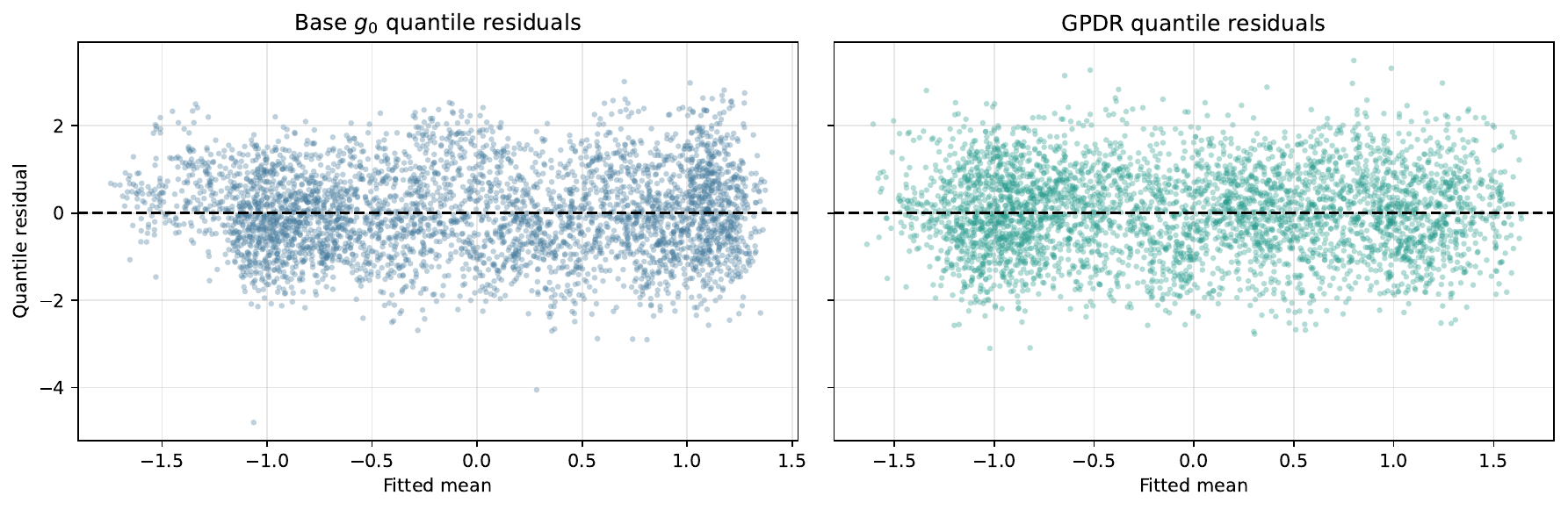}
    \caption{\small Quantile residuals for parametric model (left) and GPDR (right) versus
    estimated posterior mean values for the weather data.}
    \label{fig:weather_residuals}
\end{figure}

Figure~\ref{fig:weather_residuals} plots quantile residuals for the parametric model
and the GPDR predictive densities versus the fitted mean values.  
The quantile residual for $(y_i,x_i)$ for a predictive model is obtained by
transforming $y_i$ by the predictive cdf and then transforming
again by the inverse standard normal cdf $\Phi^{-1}(\cdot)$.    
For example, for the parametric model this is $\Phi^{-1}(G_0(y_i|x_i))$.  The
quantile residuals for GPDR are defined similarly.  
Under correct model specification the quantile residuals will be standard normal.
The figure shows that the quantile residuals for the parametric model (left) show
increasing variability with the mean, a pattern not apparent in the GPDR residuals.  
This indicates that the response variance increases with the mean, which is not
captured by the parametric model, in contrast to GPDR.

\begin{figure}
    \centering
    \begin{subfigure}{0.45\linewidth}
         \includegraphics[width=\linewidth]{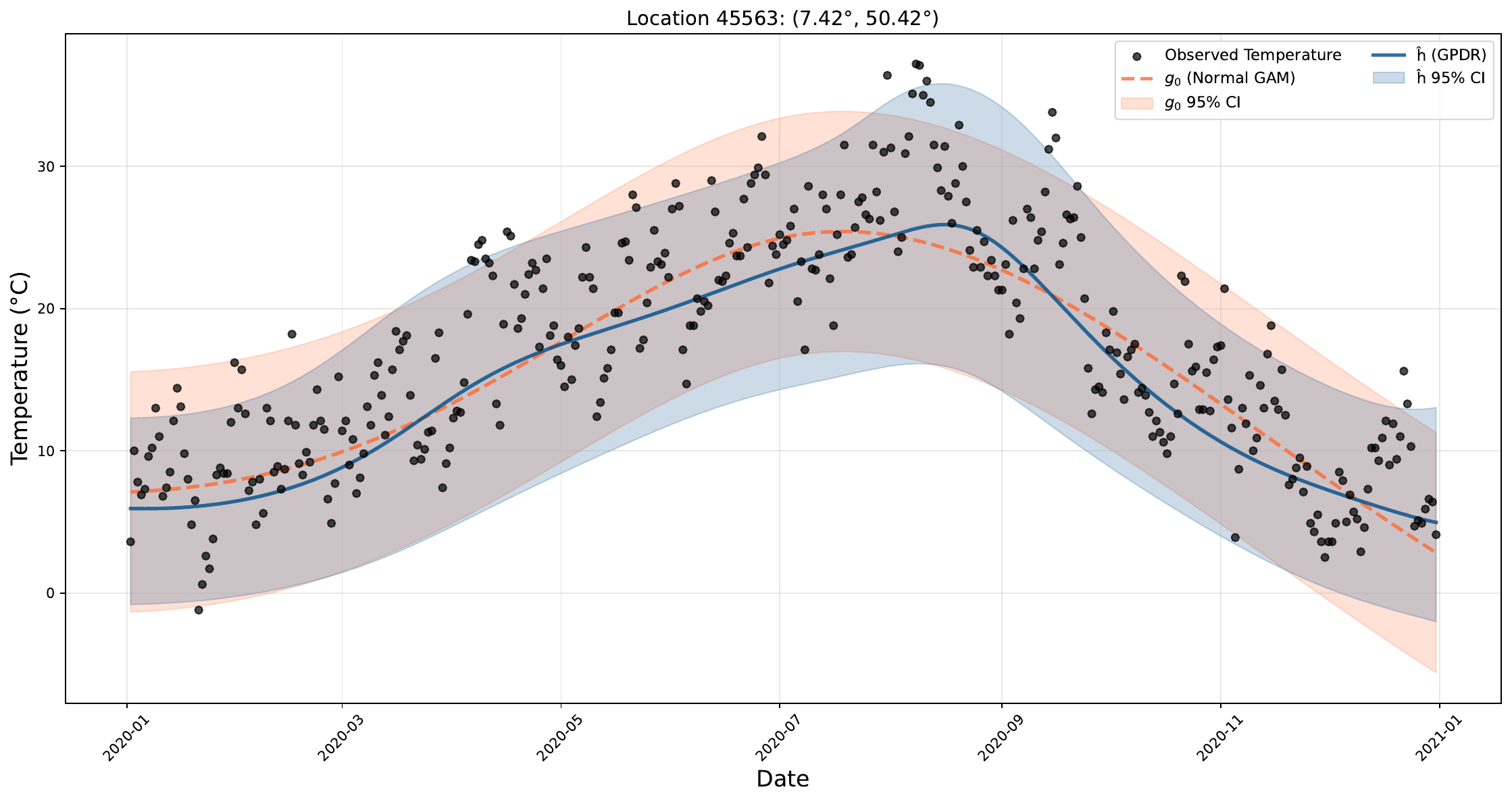}
         \caption{Temperatures at the weather station (7.42, 50.42) Andernach located in the south-west of Germany}
         \label{fig: weather time1}
    \end{subfigure}
    \hfill
    \begin{subfigure}{0.45\linewidth}
         \includegraphics[width=\linewidth]{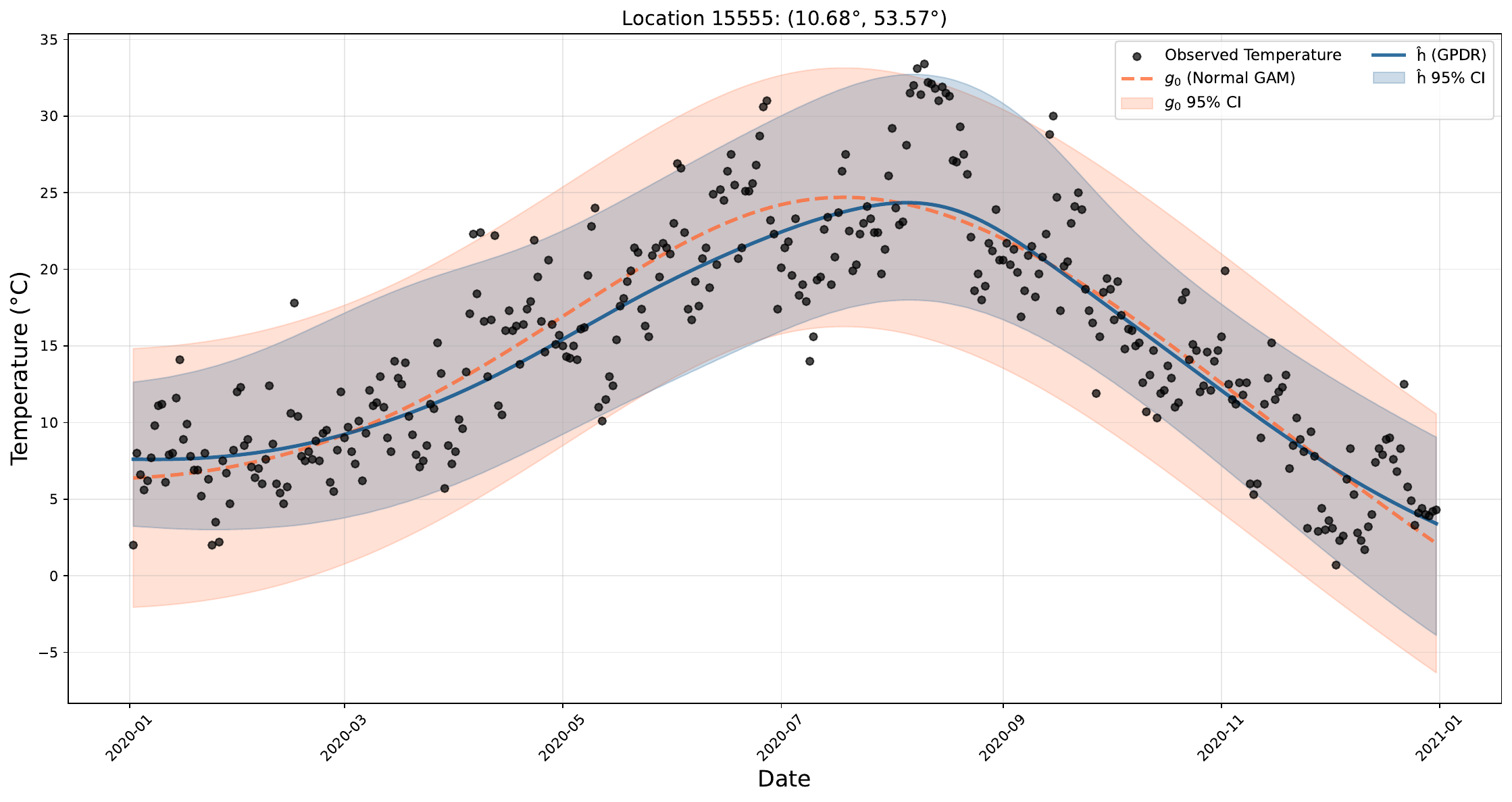}
         \caption{Temperatures at the weather station (10.68, 53.57) Grambek located in the north of Germany.}
         \label{fig: weather time2}
    \end{subfigure}

    \caption{\small Comparison of predicted temperatures over $\texttt{day}$ between two different $(\texttt{longitude},\texttt{latitude})$. The light red dashed line and shaded area denote the predictive mean and 95\% prediction intervals from $g_0(y|x)$. The light blue line and shaded area denote the predictive mean and 95\% prediction intervals from the GPDR approach.}
    \label{fig:weather_time}
\end{figure}

Figure~\ref{fig:weather_time} summarizes the predictive density as a function of time for two representative stations located in Andernach at coordinates (7.42, 50.42) and in Grambek at coordinates (10.68, 53.57) respectively. These plots are derived by fixing $\texttt{longitude}$ and $\texttt{latitude}$ while varying $\texttt{day}$. GPDR exhibits strong heteroskedasticity and results in narrower prediction intervals than for $g_0(y|x)$. Due to the additive structure, a change in location can only result in a vertical shift of the predicted mean without altering its shape over time. GPDR, however, considers the three dimensional covariate space jointly, allowing the temporal mean trajectories to vary by location and thereby accommodating more intricate spatio-temporal dynamics. 

\begin{figure}[tbh]
    \centering
    \includegraphics[width=0.9\linewidth]{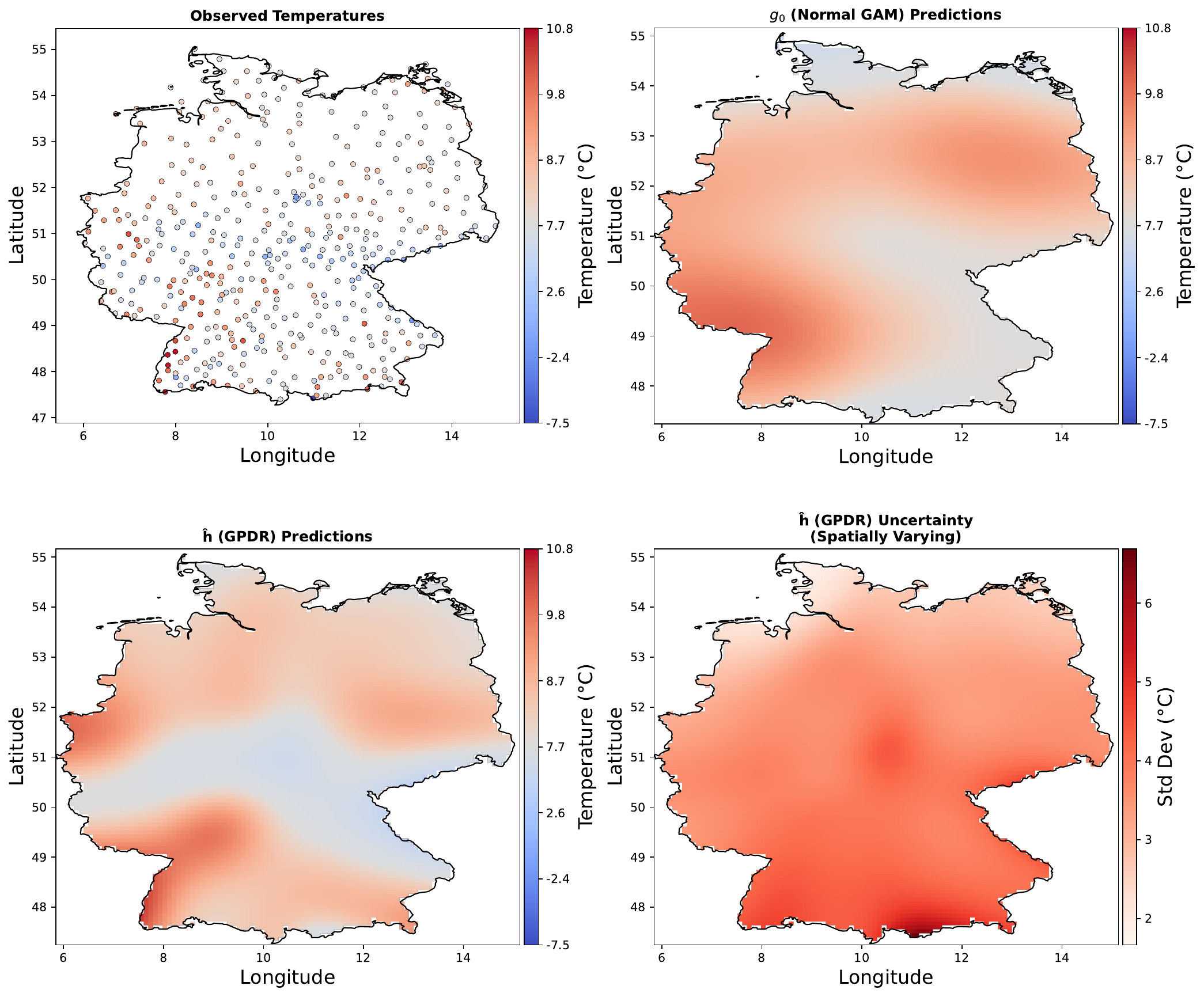}
    \caption{\small Spatial analysis given a fixed $\texttt{day}=51$. Top left panel: The observed value of temperature over locations. Top right panel: the temperature prediction from $g_0(y|x)$. Bottom left panel: the temperature prediction from GPDR. Bottom-right panel: the predictive standard deviation from GPDR. The predictive standard deviation from $g_0(y|x)$ is fixed as 4.26~°C.}
    \label{fig:weather_space}
\end{figure}

Similarly, we can examine spatial variation by fixing $\texttt{day}$ and varying $\texttt{longitude}$ and $\texttt{latitude}$. Figure~\ref{fig:weather_space} summarizes the estimated spatial effect on $\texttt{day}=51$.
The top left panel shows the observed temperatures, while top right and bottom left panels show the predicted mean, $\mathbb{E}[Y\vert x]$, under $g_0(y|x)$ and GPDR respectively, with GPDR capturing the spatial pattern of observed temperatures more accurately.
The bottom right panel reports the predictive uncertainty estimated by GPDR in terms of the conditional standard deviation. The standard deviation under the base model, $\sigma$, does not vary with the covariates, whereas GPDR identifies substantial spatial heteroscedasticity. Uncertainty is lower in northern Germany, where the terrain is flat and the proximity to the sea moderates temperature, and higher in regions with complex topography, reflecting more intricate weather phenomena that are difficult to capture without additional covariates such as elevation. 

\subsection{Gini data}\label{subsec: gini}
Finally, we illustrate GPDR for correcting a non-Gaussian parametric model. The Gini index is a widely used economic measure of income inequality within a
country \citep{gini1912variabilita}. It takes values in \([0,1]\), where 0
indicates perfect equality and 1 indicates maximal inequality. We analyze a
country-year panel constructed from the World Bank’s \emph{World Development
Indicators}, accessed via the R package \texttt{WDI} \citep{Are2025}. The dataset covers 152 countries over the period
1991--2024. Some countries do not have records for the full 34-year span resulting in $n = 2,093$ observations with training data $n_{train} = 1883$ and  test data $n_{test} = 210$.

To understand how other socioeconomic indicators are related to the Gini index, we consider a total of 5 covariates including the log gross domestic product per capita \texttt{log\_gdp}, the proportion of urban population \texttt{urban}, the unemployment rate \texttt{unemp}, the trade openness (in \%) \texttt{trade} and year of recording \texttt{year}. Estimating non-parametric conditional densities is challenging here due to the relatively small sample size compared with the number of covariates and the bounded, non-Gaussian nature of the response. 

We use a beta regression model \citep{FerCri2004,CriZei2010} as a parametric regression model for the data. In particular $y_i\vert x_i \sim\BetaD(\mu_i\phi,(1-\mu_i)\phi)$, where the mean is linked to a linear predictor via a logit transformation, $\logit(\mu_i)=\beta_0+x_i^\top\beta$, and $\phi>0$ is a precision parameter, so that $g_0(y\vert x)$ is a beta distribution. The parametric model is trained via maximum likelihood estimation. As a more diffuse version of the parametric density $g_0(y|x)$ to use as the base model density $g(y|x)$ for GPDR we consider $\BetaD(\mu_i\phi/2,(1-\mu_i)\phi/2)$.

\begin{figure}[tbh]
    \centering
    \includegraphics[width=0.9\linewidth]{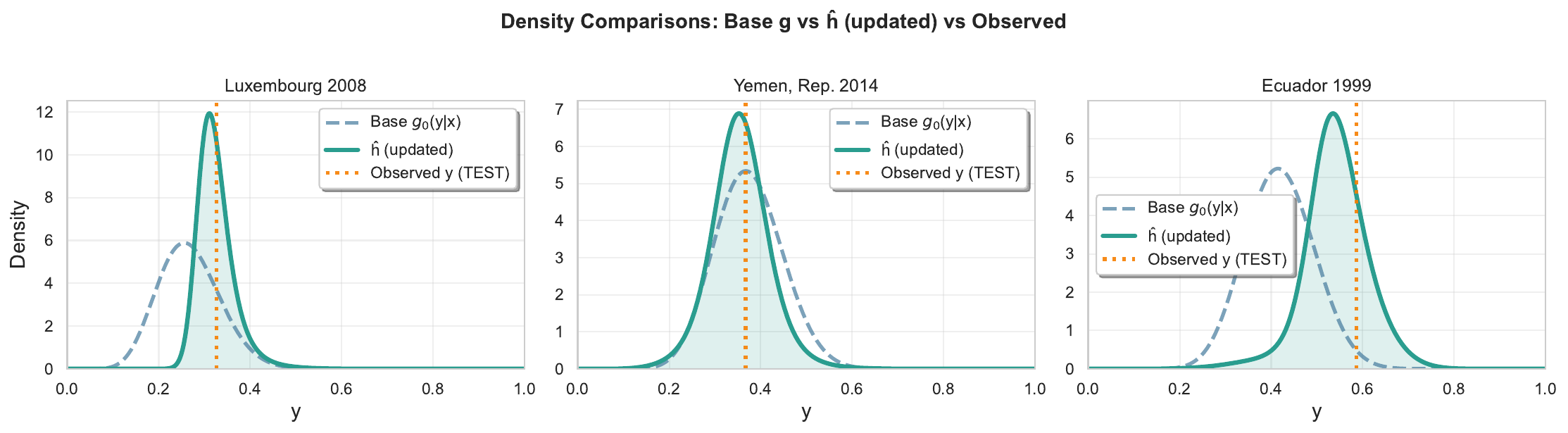}
    \caption{\small GPDR Predictive density (solid), parametric model predictive density (dashed), and observed data $y$ (dotted) for three representative data points from the test set.}
    \label{fig:gini}
\end{figure}

Figure~\ref{fig:gini} shows the estimated conditional densities under the base model and under GPDR for three selected observations. The shape for $h(y\vert x,f)$ differs with the vector of observed covariates $x$ and the estimated densities clearly deviate from $g_0(y|x)$. Table~\ref{tab:metrics} reports the log-score, the MSE, and the calibration metrics evaluated on a hold-out test set, 
demonstrating the improvement that GPDR provides over the parametric model.  

\begin{figure}
    \centering
    \includegraphics[width=0.9\linewidth]{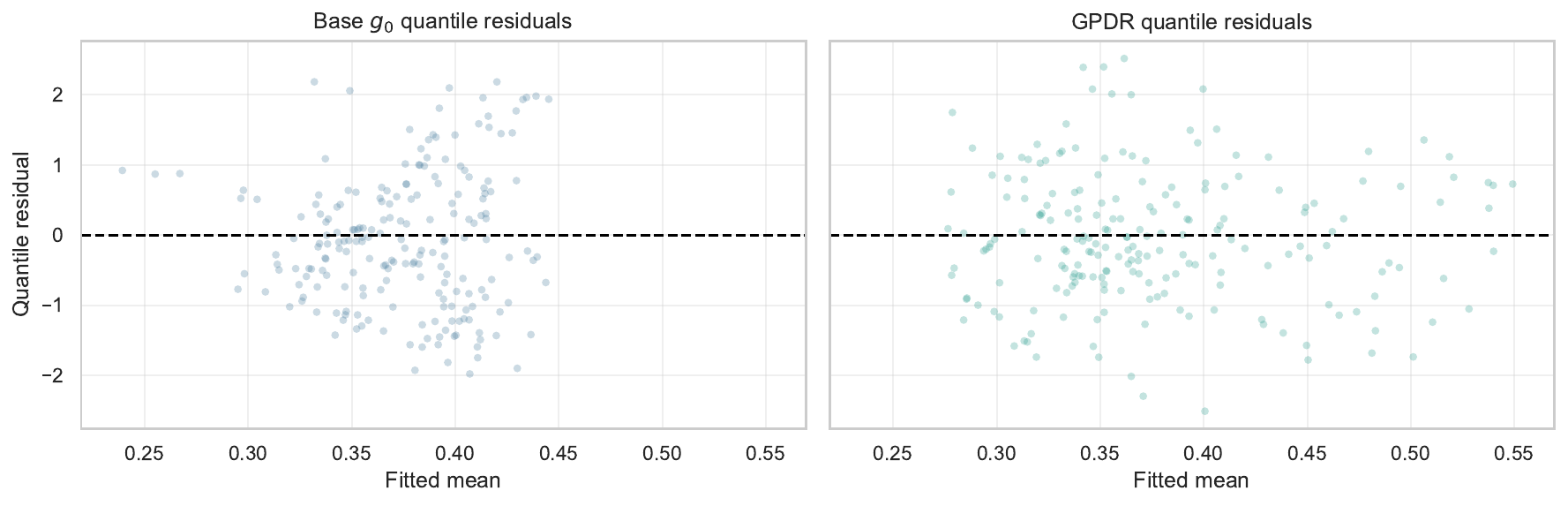}
    \caption{\small Quantile residuals for parametric model (left) and GPDR (right) versus
    estimated posterior mean values for the Gini data.}
    \label{fig:gini_residuals}
\end{figure}

Figure~\ref{fig:gini_residuals} plots the quantile residuals for the parametric model
and the GPDR predictive densities versus the fitted mean values.  
The figure shows that there is a trend in the mean level of the residuals for
the parametric model, indicating that the mean response isn't well described by
the parametric model, with GPDR providing a better fit.

\section{Discussion}\label{sec:Discussion}
This paper introduced a generalized Bayesian formulation for logistic Gaussian process density regression that replaces the intractable likelihood with a loss likelihood based on the Hyvärinen score. This loss likelihood uses only derivatives of the log-density with respect to the response, eliminating observation-specific normalizing constants that make conventional Bayesian inference for logistic GP density regression difficult. We further developed a scalable inference scheme that combines sparse inducing points with a structured variational approximation, enabling practical application on large datasets. We demonstrated improved predictive fit and calibration 
across three examples: a misspecified Gaussian regression model, a 
large spatio-temporal German weather dataset, and a bounded-response 
Gini index dataset with a beta regression base model. In each case, 
the GP correction provided interpretable model criticism, revealing 
how a parametric regression used to determine the base model 
was misspecified.

Our GPDR approach can model complex response-covariate relationships as it is not restricted by a parametric form while retaining interpretability through centering the GP prior on a parametric regression model. The GP correction acts as a transparent, data-driven adjustment revealing how an initial parametric regression is misspecified, which can be used to guide model improvement. As illustrated by the Gini data example, our approach can be used with non-Gaussian base models. 
Together, the inducing-point variational approach and the generalized 
Bayes formulation with the Hyv\"{a}rinen score scaled successfully to 
a dataset with over 150,000 observations.

We consider a Gaussian mixture variational approximation with diagonal covariance matrices and a single set of inducing points employed only for the first derivative resulting in a relatively small number of variational parameters. However, alternative approximations further exploiting the Gaussian conjugacy of the generalized posterior under the Hyvärinen score could potentially tighten the approximation error, and represent a natural direction for further research.

While GPDR performed well across all examples considered, establishing 
theoretical guarantees for Hyv\"{a}rinen score generalized posteriors 
in logistic GP density regression would be valuable. Extensions to 
high-dimensional covariates via subspace projections as in 
\cite{tokdar+zg10}, and to multivariate responses, are promising 
directions for future work.

\section*{Acknowledgements}

David Nott's research was supported by the Ministry of Education, Singapore, under the Academic Research Fund Tier 2 (MOE-T2EP20123-0009).  We thank Fran{\c{c}}ois-Xavier Briol for helpful comments.  

\FloatBarrier
\bibliographystyle{apalike}
\addcontentsline{toc}{section}{\refname}
\bibliography{logistic-GP}


\appendix
\renewcommand{\thetable}{\Alph{section}\arabic{table}}
\setcounter{table}{0}
\renewcommand{\thefigure}{\Alph{section}.\arabic{figure}}
\setcounter{figure}{0}
\renewcommand{\theequation}{\Alph{section}.\arabic{equation}}
\setcounter{equation}{0}
\renewcommand{\thesection}{Appendix \Alph{section}}

\section{Expression for terms in $\mathcal{H}(y_i,x_i,f_{.i})$}\label{app:score}

Since $\hbar(y|x,f)=\log g(y|x)+f(S(x),G(y|x))-\log c(x,f)$, where
$c(x,f)=\int_0^1 \exp(f(S(x),z))\,dz$,  we have
\begin{align*}
\hbar_y(y|x,f) & =\frac{g_y(y|x)}{g(y|x)}+f_z(v)g(y|x),
\end{align*}
where $v=(w^\top,z)^\top$, and $w=S(x)$, $z=G(y|x)$. 
Also,
\begin{align*}
  \hbar_{yy}(y|x,f) & = \frac{A(y,x)}{B(y,x)}+C(y,x,f)
\end{align*}
where
\begin{align*}
  A(y,x) & = g(y|x)g_{yy}(y|x)-g_y(y|x)^2 \\
  B(y,x) & = g(y|x)^2 \\
  C(y,x,f) & = f_z(v)g_y(y|x)+g(y|x)^2 f_{zz}(v).
\end{align*}

Substituting the above expressions for \(\hbar_y(y\mid x,f)\) and \(\hbar_{yy}(y\mid x,f)\) into \eqref{hscore}, and collecting terms, gives
$$ \mathcal H\{y,x,f_z(v),f_{zz}(v)\}=H_g(y,x) +g(y\mid x)^2f_z(v)^2
   +4g_y(y\mid x)f_z(v)+2g(y\mid x)^2f_{zz}(v)
$$
where
$$H_g(y,x):=2\frac{\partial^2}{\partial y^2}\log g(y\mid x)
  +\left\{\frac{\partial}{\partial y}\log g(y\mid x)\right\}^2=\frac{2g_{yy}(y\mid x)}{g(y\mid x)}
  -\left\{\frac{g_y(y\mid x)}{g(y\mid x)}\right\}^2
$$
is the Hyvarinen score associated with the base density \(g(y\mid x)\). The term \(H_g(y,x)\) does not depend on \(f\). Hence, as a function of the Gaussian-process derivatives, \(\mathcal H\) is quadratic in \(f_z(v)\) and linear in \(f_{zz}(v)\).

\section{Approximation of the optimization target}
\label{app:klbound}

The optimization objective defined in \eqref{opt obj} consist of two terms: a KL-divergence term involving the Gaussian mixture variational posterior and the expected Hyvärinen score. 
Let the variational posterior be given by
\begin{equation*}
    q(\widetilde{f}_z)=\sum_{k=1}^{C} \omega_k\,q_k(\widetilde{f}_z),
\qquad
    q_k(\widetilde{f}_z)=\mathcal{N}(\mu_k,\Sigma_k),
\end{equation*}
where $\{\omega_k\}_{k=1}^{C}$ are mixture weights satisfying $\sum_k \omega_k=1$.
Write $\pi(\widetilde{f}_z)=\mathcal{N}(0,K)$ for the prior, where $K=\Delta_{1,1}(\widetilde{V},\widetilde{V})$.

The KL divergence term in $\mathcal{L}(\lambda)$ is not available in closed form. \citet{dezfouli+b15} handle this issue by decomposing the negative KL term into an entropy term and a 
cross-entropy term, using a Jensen lower bound for the entropy term and evaluating the cross-entropy term exactly. Here, we consider Goldberger's approximation given in \eqref{eq:goldberg_approx}, which is cheaper to evaluate. This approximation allows us to consider separate KL-divergences for each Gaussian component $\KLD{q_k(\widetilde f_z)}
{\pi(\widetilde f_z)}$, $k=1,\dots,C$. For each Gaussian component $q_k(\widetilde{f}_z)=\phi(\widetilde{f}_z;\mu_k,\Sigma_k)$ and prior $\pi(\widetilde{f}_z)=\phi(\widetilde{f}_z;0,K)$, the divergence admits the following closed-form expression:
\begin{equation*}
    \KLD{q_k(\widetilde f_z)}
{\pi(\widetilde f_z)}
=
\frac{1}{2}
\left[
\mathrm{tr}(K^{-1}\Sigma_k)
+
\mu_k^\top K^{-1}\mu_k
-
m
+
\log|K|
-
\log|\Sigma_k|
\right],
\end{equation*}
where $m$ is the number of inducing points.

Following \ref{app:score} the Hyvärinen score is quadratic in $f_{z}$ and linear in $f_{zz}$, so that its expectation is available in closed form. Considering the plug-in estimate $\widehat{f}_{zz}=E[f_{zz}\vert \widetilde{f}_z]$ for $f_{zz}$ suggests using $E\widehat{f}_{zz}=\Delta_{2,1}(V,\widetilde{V})\Delta_{1,1}(\widetilde{V},\widetilde{V})^{-1}\widetilde{f}_z$. In the variational
optimization, we consider draws of $\widetilde{f}_z$ from the
variational distribution, which is a Gaussian mixture.  
To ensure that the sample based approximation remains differentiable
for employing reparametrization gradients, we consider a 
Gumbel softmax trick \citep{jang2016categorical,maddison2016concrete} 
to relax the mixture sampling.  Specifically, 
for each mixture component $k=1,\dots,C$, we independently draw
$\epsilon_k\sim\mathcal{N}(0,I_m)$ and
$\gamma_k\sim\mathrm{Gumbel}(0,1)$, and compute
\[
u_k=\mu_k+\Sigma_k^{1/2}\epsilon_k,
\qquad
r_k=
\frac{\exp\{(\log\omega_k+\gamma_k)/\tau\}}
{\sum_{\ell=1}^C
 \exp\{(\log\omega_\ell+\gamma_\ell)/\tau\}},
\]
where the temperature is fixed at $\tau=0.5$.
These component draws are combined using the random
Gumbel-softmax weights to form a random working observation $\widetilde{f}^\circ_z=\sum_{k=1}^C r_k u_k$, which we use to evaluate the second-derivative imputation in each step of the SGA algorithm as
\[
\widehat{f}_{zz}
=
\Delta_{2,1}(V,\widetilde{V})
\Delta_{1,1}(\widetilde{V},\widetilde{V})^{-1}
\widetilde{f}^\circ_z.
\]
Since $\widetilde{f}^\circ_z$ is a Gumbel-softmax relaxation of mixture sampling rather than an exact random draw from $q(\widetilde{f}_z)$, it permits path-wise differentiation with respect to the variational parameters.

\section{Hyperparameter selection}\label{appendix sec: hyperselection}

The implementation of GPDR requires the choice of several hyperparameters, including the covariance scale and lengthscale parameters in the Gaussian process prior, the learning weight $\beta$ in the generalized Bayesian update, the number of inducing points, and the number of Gaussian mixture components in the variational approximation. Recall from Section~\ref{sec:Logistic} that the Gaussian process covariance is written as
\[
\Delta(v,v')
=
\sigma_f^2
\exp\left(
-\sum_{j=1}^{d}\theta_{cj}^2(w_j-w_j')^2
-\theta_r(z-z')^2
\right),
\]
where $v=(w^\top,z)^\top$. We tune these hyper-parameters through the equivalent lengthscale parametrization
\[
\theta_{cj}^2=\frac{1}{2l_x^2},
\qquad
\theta_r=\frac{1}{2l_z^2},
\]
using a common covariate lengthscale $l_x$ for all transformed covariate dimensions and a separate lengthscale $l_z$ for the transformed response coordinate. Thus, $l_x$ controls smoothness in the covariate directions and $l_z$ controls smoothness in the $z$ direction.

We select the kernel scale $\sigma_f$, the lengthscales $l_x$ and $l_z$, and the learning weight $\beta$ by 3-fold cross-validation. The cross-validation search is carried out over a finite candidate set. For the kernel hyperparameters, we consider
\[
\sigma_f\in[0.01,1],\qquad
l_x\in[0.01,1],\qquad
l_z\in[0.01,1],
\]
For the learning weight, we search over
\[
\beta\in[0.01,2000],
\]
using a logarithmic scale. We first generate 100 candidates using a Latin-hypercube space-filling design over the specified ranges. This gives a sparse candidate set that still covers the search region well. Each candidate is evaluated by 3-fold cross-validation, using the average validation log-score across the three folds as the criterion. Starting from these 100 evaluated candidates, we then perform 100 additional steps of Bayesian optimization. At each step, a Gaussian process surrogate model is fitted to the previously evaluated candidates, represented in the normalized four-dimensional hyperparameter space, and a new candidate is selected by maximizing the expected improvement acquisition function over the same search region. The newly selected candidate is then evaluated using the same 3-fold cross-validation procedure and added to the set of evaluated candidates before refitting the surrogate model. The final hyperparameter is chosen as the candidate with the highest average validation log-score among all 200 evaluated candidates. For the weather data example, evaluating the validation log-score on the full held-out fold is computationally expensive, so we evaluate it on a one-third subsample of each validation fold while keeping the same training folds.

The number of inducing points and the number of Gaussian mixture components are fixed separately for each example before cross-validation, reflecting the different sample sizes and dimensions of the three datasets. The final settings used in the simulation, weather, and Gini examples are reported in Table~\ref{tab:hyper_sim}.

\begin{table}[tbh]
\centering
\begin{tabular}{lccc}
\toprule
Hyperparameter & Gaussian & Weather dynamics & Gini dada \\
& Sec. \ref{subsec:simulation} & Sec. \ref{subsec: weather} & Sec.\ref{subsec: gini} \\
\midrule
Kernel scale $\sigma_f$ & $0.9199$ & $0.1806$ & $0.5769$\\
Covariate lengthscale $l_x$ & $0.0786$ & $0.4901$& $0.5070$\\
Transformed-response lengthscale $l_z$ & $0.3363$ & $0.2360$ & $0.8355$ \\
Learning weight $\beta$ & $1031.17$ & $29.6$ & $493.61$\\
Adam initial learning rate & $0.5$& $0.5$ & $0.5$ \\
Number of inducing points & $150$& $1000$ & $200$ \\
Number of mixture components $C$ & $2$ & 2& 2\\
\bottomrule
\end{tabular}
\caption{\small Hyperparameter settings used for the empirical studies.}
\label{tab:hyper_sim}
\end{table}

\newpage 

\section{Additional simulation results}\label{appendix sec: pygam on weather}

\begingroup
\renewcommand{\thefootnote}{}
\footnotetext{\tiny *Although DDPStar provides competitive predictive metrics, worth to note that it requires much more computation time: approximately 4 hours for the simulations example and 8 hours for the weather data example, compared with at most 20 minutes for each of the other methods in these two examples.}
\endgroup

The summary of the predictive performance comparison of GPDR relative to the benchmark methods is shown in Table \ref{tab:metrics}. 

\begin{table}[ht!] 
\centering
\small
\linespread{1.2}\selectfont
\begin{tabular}{@{}lrrrr@{}}
Method & log-score ($\uparrow$) & RMSE ($\downarrow$) & coverage & avg. 95\% length\\
\hline \hline
\multicolumn{5}{l}{\textit{Gaussian regression}}\\
parametric base model & 0.2071 & 0.1965 & 0.9384 & 0.7454\\
GPDR & 0.5582 & 0.1615 & 0.9533 & 0.6061\\
GAMLSS & \textbf{0.5627} & \textbf{0.1614} & 0.9460 & 0.5833\\
TRAM & 0.5158 & 0.1620 & 0.9459 & 0.5995\\
QRF & -- & 0.1645 & 0.9041 & 0.5493\\
DDPstar* & 0.5445 & 0.1615 & 0.9495 & 0.5879\\
\hline \hline
\multicolumn{5}{l}{\textit{German weather data}}\\
parametric base model & -0.7792 & 4.2602 & 0.9551 & 16.8838\\
GPDR & \textbf{-0.6509} & \textbf{4.0354} & 0.9546 & 15.0426\\
GAMLSS & -0.8111 & 4.4943 & 0.9528 & 17.4958\\
TRAM & -0.7982 & 4.2237 & 0.9509 & 17.5773\\
QRF & -- & 6.6116 & 0.9726 & 28.4464\\
DDPstar* & -0.6586 & 4.0913 & 0.9520 & 15.0124\\
\hline \hline
\multicolumn{5}{l}{\textit{Gini index}}\\
parametric base model & 1.1577 & 0.0778 & 0.9381 & 0.2853\\
GPDR & 1.4447 & 0.0689 & 0.9333 & 0.2439\\
GAMLSS & 1.4807 & 0.0719 & 0.9286 & 0.2489\\
TRAM & 1.2845 & 0.0745 & 0.9286 & 0.2665\\
QRF & -- & \textbf{0.0577} & 0.9714 & 0.2437\\
DDPstar* & \textbf{1.5163} & 0.0756 & 0.9095 & 0.2500\\
\hline \hline
\end{tabular}
\caption{\small Comparison of performance metrics among the parametric model $g_0$, GPDR, and benchmark models over three examples. We report the average log-score per observation, RMSE, the proportion of observations lying within an equal-tailed 95\% prediction interval, and the average length of this interval. All metrics are evaluated on hold-out test data and rounded to 4 decimals. The best values for reported log-score and RMSE across each example are marked. No log-score is reported for QRF as the approach does not result in a closed form density. For GAMLSS, a normal model with additive effects consisting of a joint spatial effect and a separate effect for time is considered for the weather example, and a Beta model with separate additive effects in all covariates for the Gini example.}
\label{tab:metrics}
\end{table}

\begin{figure}[tbh]
\centering
\begin{subfigure}{0.5\linewidth}
         \includegraphics[width=1.0\linewidth,keepaspectratio]{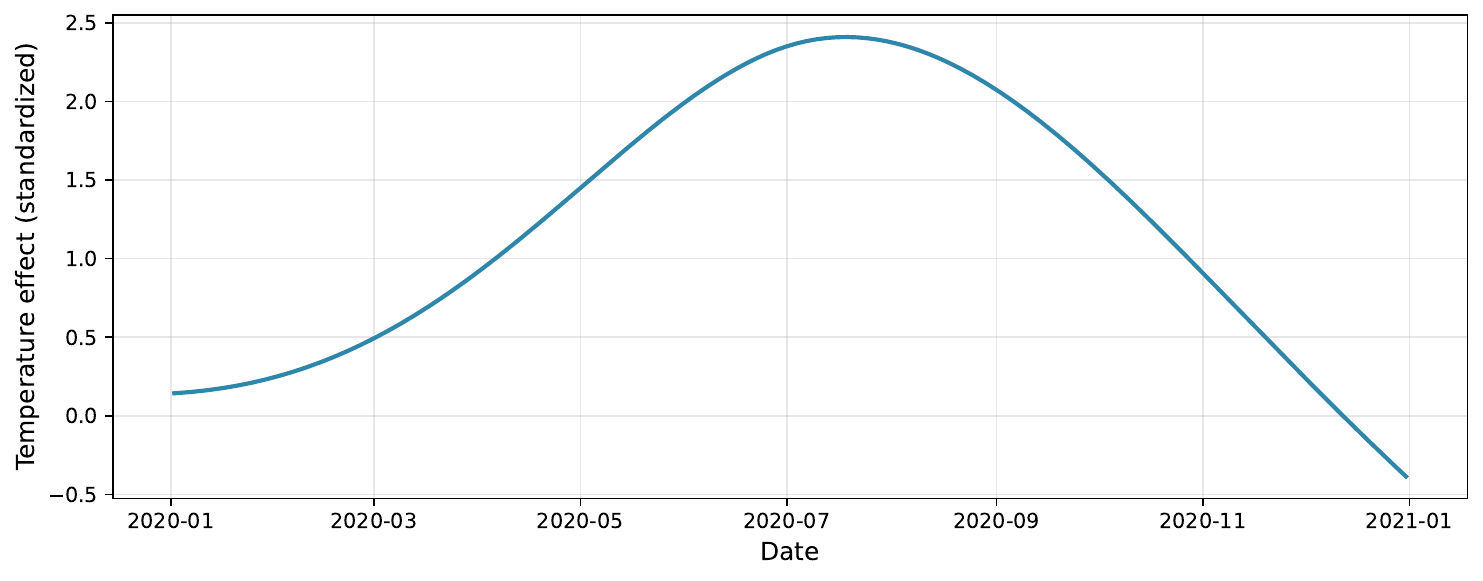}
         \caption{Temporal effect}
         \label{basemodel:time}
\end{subfigure}
\begin{subfigure}{0.5\linewidth}
         \includegraphics[width=1.0\linewidth,keepaspectratio]{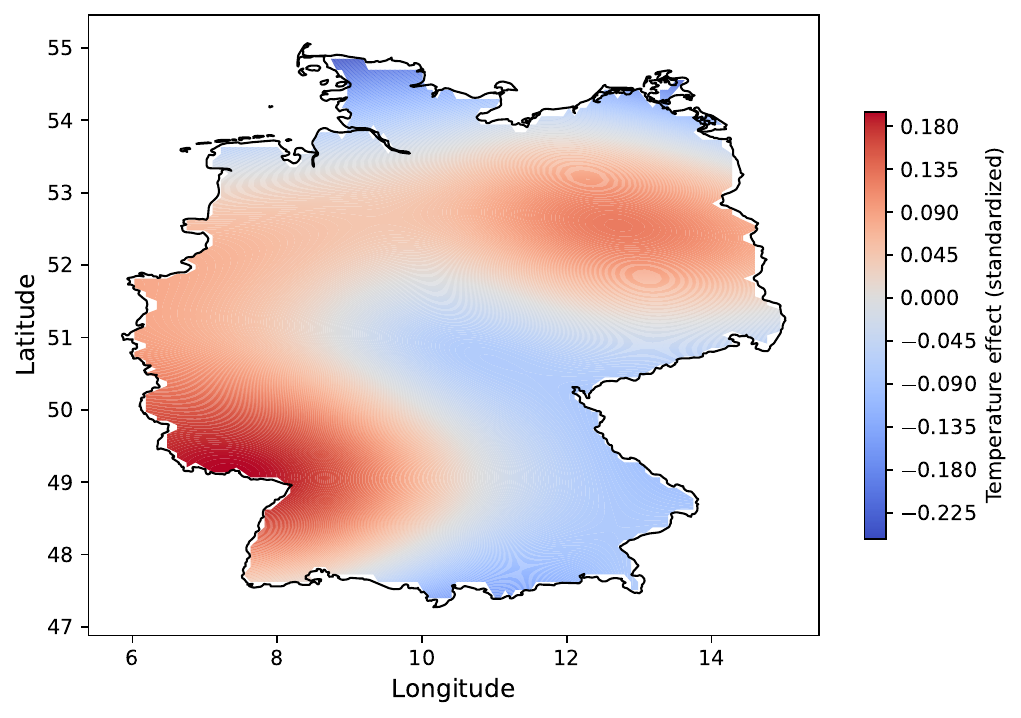}
         \caption{Spatial effect}
         \label{basemodel:space}
\end{subfigure}

\caption{\small Spatio-temporal effect estimated by base model}
\label{fig:basemodelweather}
\end{figure}

Figure~\ref{fig:basemodelweather} presents the estimated spatio-temporal effects
from the fitted base model in the weather data example described in
Section~\ref{subsec: weather}. The base model is an 
overdispersed and heavy-tailed variant of a generalized additive model
(GAM) that decomposes temperature predictions into additive components: an
intercept, a temporal effect $\mathfrak{f}_\text{temp}(\texttt{day})$, and a spatial effect $\mathfrak{f}_\text{spat}(\texttt{longitude},\texttt{latitude})$.
The plots in the figure visualize the corresponding partial dependence functions for the GAM. Panel~\ref{basemodel:time} shows how the model’s temperature predictions vary
over time (dates), averaged across all spatial locations. It is obtained from
the GAM partial dependence function for the temporal covariate and reflects
systematic temporal variation such as seasonal cycles over the observation
period. Panel~\ref{basemodel:space} displays the spatial temperature effect across
Germany after standardization. This is computed from the tensor product smooth
\(te(lon, lat)\) in the GAM, evaluated on a grid and masked to the country’s
boundaries. The diverging color map (blue--red, centered at zero) highlights
regions that are systematically cooler (blue) or warmer (red) than average,
independently of temporal effects, capturing geographic patterns such as
elevation, proximity to water bodies, and urban heat islands. Both plots use standardized temperature values (scaled to \([0,1]\)) to improve numerical stability during model fitting. 

\end{document}